\documentclass[12pt,onecolumn]{article} 
\usepackage[left=1in, right=1in, top=1in, bottom=1in]{geometry}
\usepackage{wrapfig}
\usepackage[numbers]{natbib}
\usepackage{times}
\usepackage{url}
\usepackage{graphicx}
\usepackage[usenames]{color}
\DeclareGraphicsExtensions{.pdf,.png,.jpg,.mps,.eps,.ps}
\usepackage{amsmath}
\usepackage{tablefootnote}
\def\unit #1{\,{\rm #1}}
\usepackage{cite}
\newcommand\kms{\rm \,\unit{km\,s^{-1}}}

\usepackage{graphicx}
\usepackage{multicol}

\newcommand\kev{\rm \,\unit{keV}}

\newcommand\xiunit{\rm \,erg\,cm\,s^{-1}}

\newcommand\mbh{M_{\rm BH}}

\newcommand\nh{ N_{\rm H}}

\newcommand\ks{\, \rm ks}

\newcommand\pc{\unit{pc}}

\newcommand\kpc{\unit{kpc}}

\newcommand\ev{\unit{\, eV}}

\newcommand\chandra{{\it Chandra}}

\newcommand\xmm{{\it XMM-Newton}}

\usepackage{mathptmx}

\newcommand\lion{L_{\rm ion}}

\newcommand\mout{\dot{M}_{\rm out}}

\newcommand\Ek{\dot{E}_{\rm K}}

\usepackage{times}
\usepackage{geometry}
\usepackage[utf8]{inputenc}
\usepackage{enumitem,amssymb}
\usepackage{ragged2e}
\newlist{thematic}{itemize}{8}
\setlist[thematic]{label=$\square$}
\usepackage{pifont}

\begin{document}

\huge
\begin{flushleft}
	{{\Large Astro2020 Science White Paper}\\
\medskip
\medskip
\medskip

{\LARGE \bf The physics and astrophysics of X-ray outflows from Active Galactic Nuclei.} \\
}
\end{flushleft}
\medskip
\medskip

\normalsize

\noindent \textbf{Thematic Areas:} \hspace*{60pt} $\square$ Planetary Systems \hspace*{10pt} $\square$ Star and Planet Formation \hspace*{20pt}\linebreak
$\square$ Formation and Evolution of Compact Objects \hspace*{31pt} $\square$ Cosmology and Fundamental Physics \linebreak
  $\square$  Stars and Stellar Evolution \hspace*{1pt} $\square$ Resolved Stellar Populations and their Environments \hspace*{40pt} \linebreak
  {\bf \large $\checkmark$}    Galaxy Evolution   \hspace*{45pt} $\square$             Multi-Messenger Astronomy and Astrophysics \hspace*{65pt} \linebreak

\medskip
\medskip
\medskip
\medskip
\medskip
\medskip
\medskip
\medskip

\centerline{Principal author: \textbf{Sibasish Laha}}
\centerline{University of California, San Diego}
\centerline{email: sib.laha@gmail.com, slaha@ucsd.edu}
\centerline{Phone: 858-999-4720}

\medskip
\medskip
\medskip
\medskip
\medskip

\noindent \textbf{Co-authors:} Randall Smith$^1$, Panayiotis Tzanavaris$^2$, Tim Kallman$^2$, Sylvain~Veilleux$^3$,\\
Francesco~Tombesi$^2$, Gerard~Kriss$^5$, Matteo Guainazzi$^6$, Massimo Gaspari$^7$, Jelle Kaastra$^8$,\\ 
Alex Markowitz$^{9,14}$, Mike Crenshaw$^{10}$, Ehud Behar$^{11,3}$, Keigo Fukumura$^{12}$, Anna Lia Longinotti$^{13}$, Agata Rozanska$^{14}$ 
Jacobo Ebrero$^{15}$, Gary Ferland$^{16}$, Claudio Ricci$^{17,18}$, Chris Done$^{19}$, \\
Daniel Proga$^{20}$, Mitchell Revalski$^{10}$, Andrey Vayner$^{9}$.
\medskip
\medskip
\medskip
\medskip

\noindent{\bf Institutions/affiliations:} $^1$Center for Astrophysics, Harvard, $^2$NASA Goddard Space Flight Center,
 $^3$University of Maryland, College Park, $^5$STScI, $^6$ESTEC, $^7$Princeton University, $^8$SRON,
$^9$University of California, San Diego, $^{10}$Georgia State University, $^{11}$Technion, $^{12}$James Madison University, $^{13}$INAOE-CONACYT,
$^{14}$Nicolaus Copernicus Astronomical Centre, Polish Academy of Sciences, $^{15}$ESAC, $^{16}$University of Kentucky, $^{17}$Universidad Diego Portales, $^{18}$Kavli Institute for Astronomy and Astrophysics, Peking University,
$^{19}$Durham University, $^{20}$University of Nevada, Las Vegas.

\medskip


\newpage
\noindent\textbf{Abstract:} The highly energetic outflows from Active Galactic Nuclei detected in X-rays are one of the most powerful mechanisms by which the central supermassive black hole (SMBH) interacts with the host galaxy. The last two decades of high resolution X-ray spectroscopy with \xmm{} and \chandra{} have improved our understanding of the nature of these outflowing ionized absorbers and we are now poised to take the next giant leap with higher spectral resolution and higher throughput observatories to understand the physics and impact of these outflows on the host galaxy gas. The future studies on X-ray outflows not only have the potential to unravel some of the currently outstanding puzzles in astronomy, such as the physical basis behind the $\mbh-\sigma$ relation, the cooling flow problem in intra-cluster medium (ICM), and the evolution of the quasar luminosity function across cosmic timescales, but also provide rare insights into the dynamics and nature of matter in the immediate vicinity of the SMBH. Higher spectral resolution ($\le 0.5\ev$ at $1\kev$) observations will be required to identify individual absorption lines and study the asymmetries and shifts in the line profiles revealing important information about outflow structures and their impact. Higher effective area ($\ge 1000 \rm \,cm^{2}$) will be required to study the outflows in distant quasars, particularly at the quasar peak era (redshift $1\le z\le 3$) when the AGN population was the brightest. \textbf{Thus, it is imperative that we develop next generation X-ray telescopes with high spectral resolution and high throughput for unveiling the properties and impact of highly energetic X-ray outflows. A simultaneous high resolution UV + X-ray mission will encompass the crucial AGN ionizing continuum (1~eV -- 100~keV), and also characterize the simultaneous detections of UV and X-ray outflows, which map different spatial scales along the line of sight.}  \\

\noindent\textbf{\large 1. Introduction and our current knowledge}\\

It is nowadays well accepted that most galaxies have passed through a recurrent active galactic nucleus (AGN) phase. During the active phase, the supermassive black hole (SMBH) at the center of the galaxy accretes matter ({\it feeding}) at a significant rate and thus shines brightly in most spectral bands. During this phase, the SMBH ejects energetic particle {\it outflows} in the form of jets and/or winds which interact with the host galaxy and the ambient medium ({\it feedback}), such as interstellar medium, inter-galactic medium (IGM), and intra-cluster medium (ICM) thus regulating the SMBH-galaxy co-evolution across cosmic time. The blue-shifted absorption (and sometimes emission) lines in the energy range $0.2-10\kev$, arising out of ionic transitions from several astrophysically abundant elements (C, N, O, Ne, Fe etc) are signatures of highly energetic X-ray outflows, which are a crucial form of AGN {\it feedback} mechanism. Since X-rays probe some of the most energetic environments near the SMBH, these X-ray outflows are also important sources of information for the dynamics and state of matter in the strong gravity regime of the SMBH. To accurately quantify the physical state and dynamics of these outflows we need to measure the properties of the outflowing plasma/gas, such as the 1. ionization parameter ($\xi$), 2. equivalent neutral hydrogen column density ($\nh$), 3. velocity ($v$), 4. electron density ($n_{\rm e}$), 5. distance ($r$) of these ionized absorbers from the SMBH and 6. covering factor of the outflows. \\

\noindent\textbf{\large 1.1 Advances in the last two decades of high resolution X-ray spectroscopy}\\

In the last two decades of high resolution X-ray spectroscopy with \chandra{} and \xmm{} we have greatly improved our understanding of X-ray ionized outflows. Several sample studies of local Seyfert galaxies have found that the warm absorbers (WA) are present in almost $\sim 70\%$ of the nearby AGN and they have an ionization parameter in the range $\log\xi/\xiunit=-1$ to $3$, column density in the range $\log\nh/\rm cm^{-2}= 20-22$ and outflow velocities of $v=100-2000 \kms$ \citep[][]{2005A&A...431..111B,2014MNRAS.441.2613L}. Detailed X-ray spectroscopic studies on several bright nearby sources such as MCG-6-30-15, IRAS~13349+2438, NGC~4051 have revealed stunning spectral signatures and multi-zone outflowing absorbers hinting at complex acceleration mechanisms at work, which need to be revealed more clearly with next generation instruments. The higher velocity outflows (with $v\sim 10,000-30,000 \kms$) known as the ultra-fast outflows (UFOs) on the other hand, have been detected up to $\sim 40\%$ of the local AGN with extreme physical parameters such as $\log(\xi/\xiunit)=3-5$ and column density in the range $\log(\nh/\rm cm^{-2})=22-24$ \citep{2011ApJ...742...44T,2012MNRAS.422L...1T}. Below we summarize the current status on the most prominent issues of X-ray ionized outflows:\\

\begin{wrapfigure}{r}{0.5\textwidth}
\begin{center}	

\includegraphics[width=5.5cm,angle=-90]{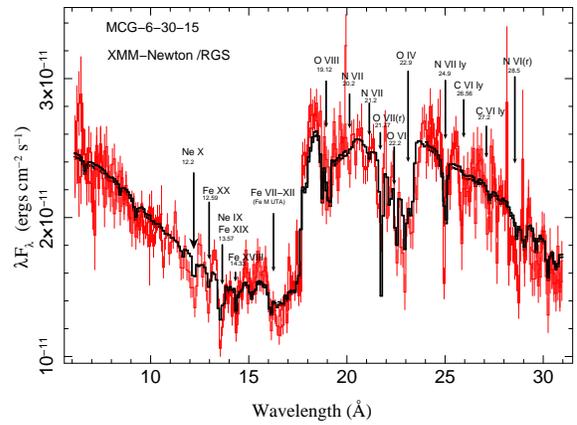}

	\caption{ \small \it The X-ray warm absorber features revealed in the high resolution \xmm{} RGS spectra of MCG-6-30-15. \citep{2014MNRAS.441.2613L}.  }
 \label{Fig:MCG6}
\end{center}
\end{wrapfigure}

(a) {\it Presence of dust}: In several Seyfert galaxies such as MCG-6-30-15 and IRAS~13349+2438, the warm absorbers have been found to be associated with dust \citep{2001ApJ...554L..13L,2013MNRAS.430.2650L}. For MCG-6-30-15 (see Figure \ref{Fig:MCG6}), the dust column density measured in X-rays agrees with that obtained from optical reddening studies, confirming the presence of dust in the X-ray outflows (dust embedded in partially ionized material). From a kinematical point of view, it has been theoretically and observationally shown that dust enhances the radiation efficiency in driving the outflows \citep{2008MNRAS.385L..43F,2013MNRAS.431.3127V,2018MNRAS.476..512I} and hence plays an important role in accelerating the outflows, even for sub-Eddington sources.

\smallskip

(b) {\it The unresolved transition array (UTA) of Fe M shell}: A deep absorption trough is detected in some of the X-ray spectra of AGN exhibiting WA in the wavelength range $15-17\rm \AA$, which is commonly known as the Fe M-shell UTA absorption, which arises due to the $2p-3d$ inner-shell absorption by iron M-shell ions \citep{2001A&A...365L.168S}. These features have been suggested to carry a large fraction of the mass outflow rates in the warm absorbers \citep{2014MNRAS.441.2613L} and hence serve as an important tool for feedback.

\smallskip

(c) {\it Asymmetry in absorption features:} One of the most detailed high spectral resolution views of a warm absorber has been obtained for NGC ~3783 using a $900\ks$ \chandra{} exposure \citep{2002ApJ...574..643K}. The spectrum revealed absorption lines from {H and He like ions of N, O, Ne, Mg, Al, Si and S, and most absorption lines showed asymmetry having more extended blue wings than red wings}. These may be due to an asymmetric outflow geometry or a systematic velocity gradient of the absorbers along our line of sight, intercepting different (kinematic) parts of a continuous outflow.

\smallskip

	(d) {\it The gap in the ionization parameter:} Studies of WA in samples of local Seyfert galaxies by \citet{2014MNRAS.441.2613L} and \citet{2007MNRAS.379.1359M} have found a gap in the ionization parameter in the range $\log\xi=0.5-1.5$, which could be attributed to unstable clouds in these ionization states, which are short lived and hence not detectable \citep{2015ApJ...815...83A,2018arXiv181205154A}, or have very weak and narrow signatures which are not resolved by current instruments. \citet{2007ApJ...663..799H} and \citet{2009ApJ...703.1346B} have also found similar gaps in the ionization parameter (bimodal distribution of $\xi$) using dedicated studies of individual sources.

\smallskip

\begin{wrapfigure}[20]{r}{3.8 in}
\begin{center}	
\includegraphics[width=7.5cm,angle=0]{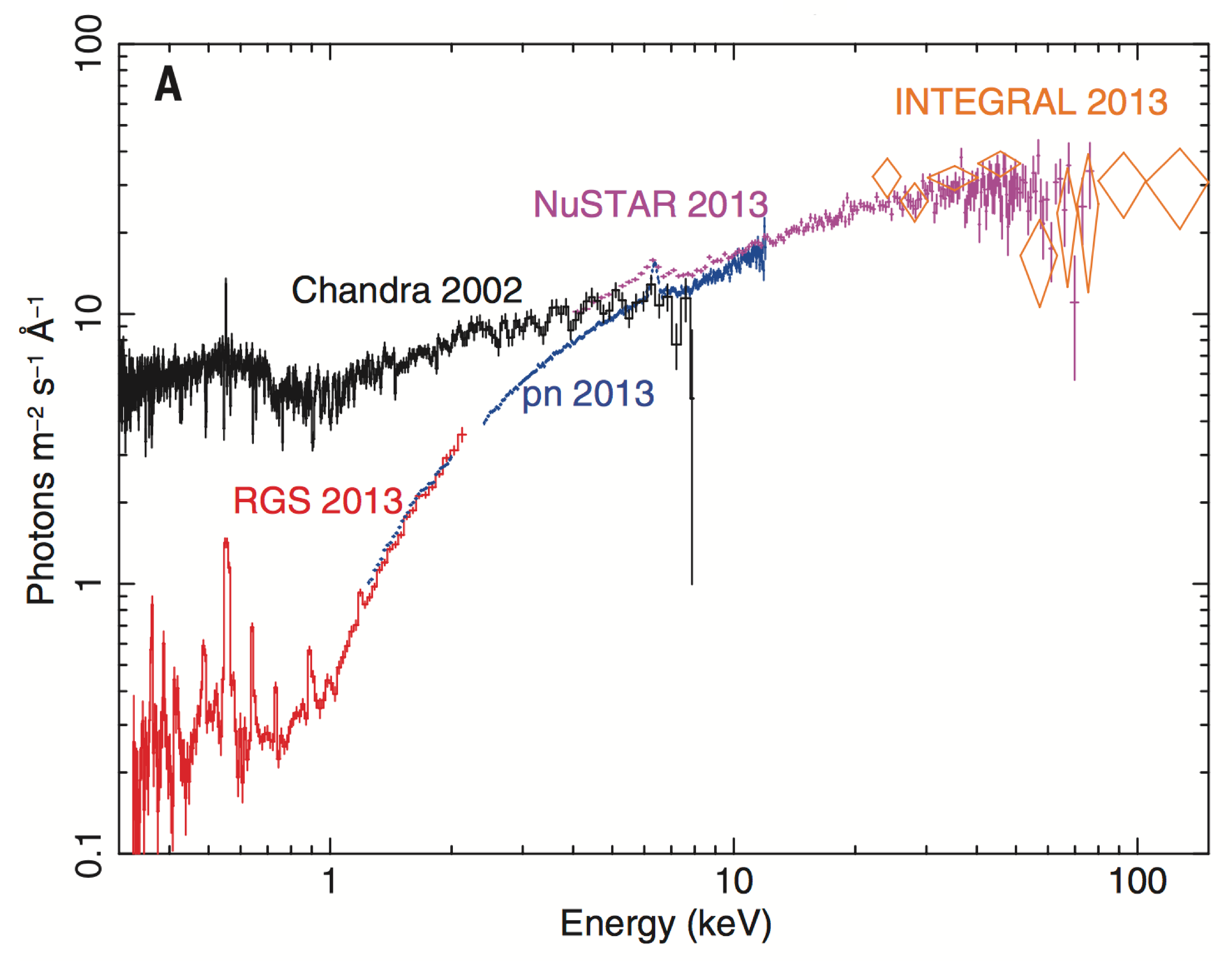}

	\caption{ \small \it The absorbed and unabsorbed X-ray spectra of NGC~5548 demonstrating the effect of the transition of variable obscuring clouds along the line of sight. The cloud obscures almost $\sim 90\%$ of the source and are detected as broad absorption line troughs simultaneously in X-rays and UV. Courtesy: Kaastra et al. (2014) \citep{2014Sci...345...64K}.  }
 \label{Fig:kaastra}
\end{center}
\end{wrapfigure}

	(e) {\it Outflows in other wavelength-bands and their connection with X-ray outflows}: There have been several attempts to understand if the UV and the X-ray outflows are similar in nature and whether they exist in multiphase outflowing gas/plasma (e.g., the unified absorber scenario by \citet{1998ApJ...503L..23M}). However, the current insufficient spectral resolution in the X-rays does not allow us to make a direct comparison with the UV. Several large and deep campaigns of bright nearby AGN such as Mrk~509, NGC~5548, NGC~7469, NGC~3783 have revealed the presence of multiphase outflows \citep{2002ApJ...574..643K,2013MNRAS.435.3028E,2012A&A...539A.117K}.

\smallskip

	(f) {\it The time variable `eclipsing' X-ray and UV outflows}: Obscuring outflows produced by ``eclipsing clouds” orbiting at broad line region (BLR) scale have been detected in the X-ray and UV spectra of several Seyfert Galaxies undergoing significant flux and spectral variations. Their outflow velocity is half way between warm absorbers and UFO. Figure \ref{Fig:kaastra} shows the line-of-sight-obscuration variability in X-rays for the source NGC~5548 due to the passage of a thick `eclipsing cloud' . Such events have also been observed in Mrk~335 \citep{2013ApJ...766..104L},  NGC~5548 \citep{2014Sci...345...64K}, NGC~985 \citep{2016A&A...586A..72E}, NGC~3783 \citep{2017A&A...607A..28M,2018ApJ...853..166K}.

\medskip

\noindent\textbf{\large 2. The key questions to be answered in the next decade.} 
\smallskip

Below we discuss the key science questions, along with the observational requirements and the key instrument parameters required to achieve the observational aims.

\medskip

\noindent\textbf{(1)} \textbf{Where do the X-ray ionized outflows originate?}

\smallskip
The origin of the X-ray outflows is still debated and the answer very much depends on the knowledge of the location where these absorbers originate. Several competing theories propose different regions of origin based on the physical nature of the outflows. For example, the highly ionized high velocity outflows are possibly launched at the inner accretion disk where the radiation is intense, while the lower ionization slower components originate further out in the accretion disk or in the broad line region (BLR) or the torus. An accurate estimate of the distance $r=(\frac{\lion}{n_{\rm e}\xi})^{1/2}$ of the ionized outflows in a statistical sample of sources is necessary to understand their origin, where $\lion$, $n_{\rm e}$ and $\xi$ are the ionizing luminosity (integrated over $13.6\ev-13.6\kev$), electron density and ionization parameter of the plasma respectively. 

\smallskip

{\it Observational aim:} Constraining the location at which the wind is launched from the AGN, for a statistical sample of AGN in the local Universe, by estimating $\lion$, $n_{\rm e}$ and $\xi$. 

{\it Key parameters required:} High spectral resolution, of $\le 0.5 \ev$ at $1\kev$ to detect individual de-blended absorption lines to estimate $n_{\rm e}$ and $\xi$. Also a simultaneous view of UV to X-ray ($\sim 1\ev-100\kev$) spectral energy distribution (SED) is needed.\\

\noindent\textbf{(2)} \textbf{What powers the X-ray outflows?}

\smallskip

The acceleration mechanism of X-ray outflows is still an open question. For highly ionized, high velocity absorbers, where the atoms have been stripped of most of their electrons, it has been proposed that magneto-hydrodynamic processes help to accelerate the plasma \citep{1982MNRAS.199..883B,2010ApJ...715..636F}. On the other hand, for the low ionized clouds, with sufficient opacity to the incident radiation, a radiatively driven scenario is more plausible \citep{2004ApJ...616..688P}. In this scenario the presence of dust also plays a crucial role, because the presence of dust enhances the thrust due to radiation and thus helps to acclerate clouds even for sub-Eddington accreting sources. Thermally driven winds have been postulated to arise when the central AGN irradiates the accretion disk and heats it up to a larger height \citep{1983ApJ...271...70B,2001ApJ...561..684K}.

\smallskip

{\it Observational aim:} Each acceleration scenario makes specific predictions. The individual absorption lines need to be fit as per the predictions from these theoretical models. For e.g., \citet{2018ApJ...853...40F} detected MHD driven winds in the source NGC~3783 with a long 900 ks exposure with \chandra{} HETG. In addition, response of the cloud parameters such as $\xi$, $\nh$ and $v$ to the changes in the continuum flux can also provide insights into its acceleration mechanism \citep{2007ApJ...659.1022K}. 

{\it Key parameters required:}The spectral resolution should be $\le 0.5 \ev$ at $1\kev$, along with high effective area ($\ge 1,000\, \rm cm^{2}$) to fit individual line profiles. \\

\noindent\textbf{(3)} \textbf{What is the density and density-profile of the outflows?}
\smallskip

Both the density and the density profile ($n_{\rm e}\propto r^{- \alpha}$) are important quantities in estimating the mass outflow rate ($\dot{M}_{{out}}$=$4\pi\mu \, r \, N_{\rm H}v_{\rm out}m_pC_f$) and kinetic luminosity ($\Ek=\mout\times v_{\rm out}^2$) of the outflows, where $m_{\rm p}$ and $C_{\rm f}$ are the mass of proton and covering factor of the outflows respectively. Both $\mout$ and $\Ek$ measures the impact of an AGN on the host galaxy gas. The density $n_{\rm e}$ gives us a measure of the distance $r$ (see Section 2.1) and the density profile constrains the distribution of matter along the outflow. The density also serves as an important indicator of the origin of the outflow, as different regions in the vicinity of the AGN have different density.
\smallskip

{\it Observational aim:} To calculate the density of the clouds from the density diagnostic `metastable' absorption line ratios \citep{2017A&A...607A.100M}. Also, the density can be estimated using the recombination timescales of ionized clouds, as obtained from the response of the ionization parameter to changes in the source luminosity.

{\it Key parameters required:}The spectral resolution of $\le 0.5 \ev$ at $1\kev$ is necessary. For e.g., the density diagnostic metastable and ground state transitions for Si IX are separated by $\sim 0.05\rm \AA$ at $55 \rm\AA$, which requires a resolving power of at least $R\sim 1100$ \citep{2017A&A...607A.100M}. Next generation X-ray telescopes {\it Athena} and {\it Lynx} will be able to achieve these qualifications. Also a simultaneous view of UV to X-ray ($1\ev-100\kev$) spectral energy distribution (SED) is necessary for an accurate characterization of the ionization parameter of the absorbing ionized gas. \\


\noindent\textbf{(4)} \textbf{What is the link between the AGN accretion rate and the outflow kinematics (accretion-ejection connection)?}

A recent study by \citet{2017Natur.549..488R} has shown how the environments close to an SMBH are shaped due to the feedback from the central AGN, indicating a clear connection between feeding and feedback. The X-ray outflows originate in close proximity to the SMBH and are thus the most probable cause for blowing away material from the vicinity of the SMBH, thereby quenching the accretion. Moreover recent theoretical studies \citep{2013ApJ...767..156M,2017ApJ...837..149G} have predicted the presence of multi-phase gas in the vicinity of the AGN ($10-1000\pc$) which are both signatures of feeding and feedback. The hot gas (detected in X-rays) is possibly a result of intense AGN feedback while the co-spatial cooler gas (detected in radio and IR) is likely the result of hot gas condensation, which later rains toward the AGN inner accretion region via Chaotic Cold Accretion \citep{2017MNRAS.466..677G}, thus feeding the SMBH. 

\smallskip

{\it Observational aim:} Detecting inflows and outflows in spatially resolved multi-waveband observations in AGN.

{\it Key parameters required:}X-ray high spectral and spatial resolution. X-ray spatial resolution comparable to that of optical and mm-wave studies (a few milli-arc-seconds) will be ideal to detect and characterize the multi-phase gas.\\

\noindent\textbf{(5)} \textbf{How do the X-ray outflows drive large scale UV and molecular outflows and impact the host galaxy?}
\smallskip

The AGN are possibly one of the main drivers behind the $\kpc$ scale molecular outflows \citep{2017A&A...601A.143F,2018ApJ...868...10L,2014A&A...562A..21C,2013ApJ...776...27V}. These large scale molecular outflows carry large amount of kinetic luminosity and mass outflow rate and are capable of removing materials from the host galaxy and also regulate the star formation rate. However, the exact mechanism by which the AGN interacts with the molecular gas in the host galaxy and drives the outflows is not known, and X-ray outflows are a very important candidate for such an interaction \citep{2015Natur.519..436T,2017A&A...601A.143F,2018ApJ...867L..11L}. In addition, as discussed earlier, the origin and acceleration of the  multi-phase outflows from AGN detected in X-rays and UV are not yet known. Several studies have examined the correlation and coupling between the UV and X-ray outflows, but there hasn't been any consensus as to whether they are a part of the same outflow, or X-ray outflows drive the UV outflow.

\smallskip

{\it Observational aim:} A simultaneous UV + X-ray spectral view is required to characterize the broad band SED from AGN and simultaneous detection of UV outflows with X-ray outflows which probe different spatial regions. In addition, spatially resolved mm-wave and IR spectroscopic studies need to be carried out to detect outflows at different distance scales and estimate their driving mechanism and impact on the host galaxy.

{\it Key parameters required:} UV + X-ray simultaneous high resolution spectral view. In addition mm-wave probe of the host galaxy molecular gas for feeding and feedback.\\

\noindent\textbf{(6)} \textbf{Do X-ray ionized outflows play an important role in the evolution of the quasar luminosity function? }

We still do not know the reason behind the quasar luminosity function evolution. It is postulated that the central AGN at a certain stage of its accretion grows bright enough to self-quench itself by removing the fuel (gas and dust) from its vicinity, through radiation and outflows. It is therefore imperative to study statistically the presence of X-ray outflows in quasars across cosmic time (z=0-2), to investigate the role of feedback in quasar luminosity function evolution.

\smallskip

{\it Observational aim:}To study the high resolution X-ray and UV spectra of sizeable samples of AGN at different redshift ranges, to detect and characterize X-ray and UV ionized outflows. 

{\it Key parameters required:}The throughput of the X-ray telescopes should be larger by several factors than the existing \xmm{}-RGS, i.e $\ge 1000\, \rm cm^2$. This is to ensure enough flux for bright sources at $z=1-2$, the quasar peak era.\\









\newpage
{
\bibliographystyle{mn2e}
\bibliography{mybib}
}

\end{document}